# Universal behavior of the bosonic metallic ground state in a two-dimensional superconductor

Zhuoyu Chen[1,2,3*], Bai Yang Wang[1,4], Adrian G. Swartz[1,2,3], Hyeok Yoon[1,2], Yasuyuki Hikita[1,3], Srinivas Raghu[1,3,4], Harold Y. Hwang[1,2,3]

[1]*Geballe Laboratory for Advanced Materials, Stanford University, Stanford, CA 94305, USA.*

[2]*Department of Applied Physics, Stanford University, Stanford, CA 94305, USA.*

[3]*Stanford Institute for Materials and Energy Sciences, SLAC National Accelerator Laboratory, Menlo Park, CA 94025, USA.*

[4]*Department of Physics, Stanford University, Stanford, CA 94305, USA.*

* zychen@stanford.edu

**Anomalous metallic behavior, marked by a saturating finite resistivity much lower than the Drude estimate, has been observed in a wide range of two-dimensional superconductors. Utilizing the electrostatically gated $LaAlO_3/SrTiO_3$ interface as a versatile platform for superconductor-metal quantum phase transitions, we probe variations in the gate, magnetic field, and temperature to construct a phase diagram crossing from superconductor, anomalous metal, vortex liquid, to Drude metal states, combining longitudinal and Hall resistivity measurements. We find that the anomalous metal phases induced by gating and magnetic field, although differing in symmetry, are connected in the phase diagram and exhibit similar magnetic field response approaching zero temperature. Namely, within a finite regime of the anomalous metal state, the longitudinal resistivity linearly depends on field while the Hall resistivity diminishes, indicating an emergent particle-hole symmetry. The universal behavior highlights the uniqueness of the quantum bosonic metallic state, distinct from bosonic insulators and vortex liquids.**

## Introduction

The anomalous metallic state observed in various two-dimensional (2D) superconductors[1], including the LaAlO$_3$/SrTiO$_3$ interface[2–5], has attracted attention recently. In the standard paradigm for weakly interacting, disordered electronic systems, a 2D system without spin-orbit coupling cannot have zero temperature metallic phases[6,7]. The implication for thin film superconductors is that the only admissible ground states are superconductors and insulators: metallic phases and their associated transitions are prohibited, and any experimental observations to the contrary have been deemed anomalous. This viewpoint has persisted for over four decades, despite the extended history of the observation of metallic behavior approaching zero temperature in 2D superconductors[8–13]. Recent observations of the metallic state in layered materials like ZrNCl (ref. 14), MoS$_2$ (ref. 15), TiSe$_2$ (ref. 16), WTe$_2$ (ref. 17,18), the oxide interface LaAlO$_3$/SrTiO$_3$ (ref. 2–5), cuprate thin films[19,20], and artificial composite systems[21–23], have substantially expanded the family of materials hosting such anomalous metallic ground states. These studies, conducted in different ranges of temperatures (from ~ 20 mK to > 10 K), also suggest the existence of a 2D quantum superconductor-metal phase transition (QSMT)[1]. For instance, magnetoresistance oscillations in nanoporous YBa$_2$Cu$_3$O$_7$ (YBCO) thin films unambiguously show that conduction in the anomalous metal phase occurs with 2$e$ charge carriers (where $e$ is the electron charge)[20], consistent with vanishing Hall resistivity indicating particle-hole symmetry[24]. In the previous studies, the anomalous metals were observed either by applying magnetic field or by varying the carrier density/disorder of the system. Since the underlying symmetries are distinct in these cases, a key open issue is the extent to which the anomalous metals in both cases (with/without time-reversal symmetry) are similar. Here, using the gated LaAlO$_3$/SrTiO$_3$ superconducting interface, we control and investigate the QSMT combining three separate parameters: gate, magnetic field, and temperature.

## Results

**Sample preparation**

We utilize gold top-gated LaAlO$_3$/SrTiO$_3$ Hall bar devices, with structure shown in Figure 1**a**. Nine unit cells of LaAlO$_3$ is epitaxially grown on top of a 0.5 mm thick SrTiO$_3$ substrate after patterning AlO$_x$ hard mask, defining a 0.4 mm wide conducting channel. The polar nature of the wide-bandgap LaAlO$_3$ leads to a conductive interface with mobile electrons occupying the conduction band of SrTiO$_3$, confined against the interface[25] (Figure 1**b**). The gold top gate is then deposited on top of the LaAlO$_3$. Gate voltage ($V_G$) controls the depth of quantum well at the interface, and thus tunes the interfacial carrier density[26] (details about the electronic filtering and thermal anchoring are given in the Methods). Figure 1**c** shows the top gate modulation of carrier density and mobility at temperature $T$ = 5 K. With increasing $V_G$, carrier density increases, and mobility decreases due to the enhanced scattering by stronger confinement of the electron wavefunctions against the interface. Therefore, $V_G$ simultaneously modulates carrier density and disorder, both of which are relevant for 2D superconductivity. Based on the mobility and density values, we estimate that $k_F l$ is in the range of 80 ~ 200 ≫ 1 in our system (here $k_F$ and $l$ are the Fermi momentum and electron mean free path, respectively), and thus we ignore localization effects and treat the normal state of our sample as a Drude metal (i.e. the sample dimension is far below the localization length)[1].

**Gate voltage and temperature dependence**

Upon further decreasing $T$ below 0.3 K, the interface turns superconducting. Figure 1**d** shows the resistivity-versus-temperature ($R$-$T$) measurements with $V_G$ being the tuning parameter. We first focus on the lowest temperatures. At high $V_G$, $R$ vanishes and the $R$-$T$ curve behavior near zero $R$ can be described by a Berezinskii-Kosterlitz-Thouless (BKT) functional form $R = A \cdot \exp[-b/(T-T_C)^{1/2}]$, where $A$, $b$ are constants and $T_C$ is the critical transition temperature. We fit the $R$-$T$ curves and extract $T_C$. As $V_G$ decreases, the $R$-$T$ curves saturate at finite values without going to zero when $T$ approaches zero, indicating a metallic ground state. At higher temperatures, all $R$-$T$ curves exhibit an onset of resistivity drop from the normal state, in which we denote the onset temperature as $T_P$. $T_P$ represents the characteristic temperature scale for Cooper pairing[5]. We further note that for some gate voltages there are kinks in the $R$-$T$ curves lower than $T_P$ and before resistivity vanishes/saturates. We denote the characteristic temperature scale at which the kinks occur as $T_F$, and discuss their physical interpretation below. $T_P$ and $T_F$ are quantitatively defined by the peaks in the second derivative of $R$-$T$ curves. Figure 1**e** shows $T_C$, $T_F$, and $T_P$ as a function of $V_G$. Note that $T_C$ vanishes at $V_G = V_C$ while $T_P$ is still finite, suggesting the existence of a QSMT critical point, on both sides of which the conduction is dominated by Cooper pairs.

**Gate voltage and magnetic field dependence**

We measure the low-field magnetoresistance (field-induced resistivity difference) at base temperature near the critical point of the gate-tuned QSMT. Figure 2**a** shows resistivity versus magnetic field near zero field while tuning $V_G$. We observe at the low field limit 1) zero magnetoresistance within the superconducting regime, and 2) positive and linear magnetoresistance in the metallic regime. Importantly, upon leaving the superconducting phase, the onset of positive magnetoresistance is much more pronounced than the onset of zero-field resistivity near the critical point. At zero field, the slope for magnetoresistance $dR/dB$ is discontinuous within our measurement resolution down to ~ 0.4 G, more than $10^3$ times lower than the typical upper critical field $H_{C2}$ of the system. The discontinuous slope likely indicates the existence of a singular point in magnetoresistance at zero field in the anomalous metal phase, distinct from the quadratic behavior for a normal Drude metal. As a comparison, we fix the $V_G$ at 1.8 V and measure magnetoresistance with varying temperature shown in Figure 2**b**. We observe qualitatively the same onset of positive linear magnetoresistance at a certain temperature, and the low-field magnetoresistance vanishes again at higher temperatures. Specifically, the singular behavior is still pronounced for 180 mK and 200 mK, but it becomes rounded for 220 mK. To quantify these observations, $dR/dB$ as functions of $V_G$ and $T$ are plotted in Figures 2**c** and 2**d**, respectively. For the gate-tuned case, $dR/dB$ rises sharply from zero to a finite value at a critical voltage, which matches the critical voltage $V_C$ extracted from $R$-$T$ analysis within our data resolution. For the temperature-tuned case, $dR/dB$ is zero at low temperatures and becomes positive at around $T_C$ (extracted from $R$-$T$ analysis). After it peaks and drops, $dR/dB$ returns to zero at around $T_P$ (obtained from $R$-$T$ analysis). These results demonstrate that the positive linear low-field magnetoresistance is a sensitive indicator for both gate-tuned QSMT and temperature-tuned SMT.

**Magnetic field and temperature dependence**

The QSMT can also be induced by magnetic field. Figure 3**a** and Figure 3**b** plot typical magnetoresistance and magnetoconductivity curves for a superconducting sample (vanishing resistance at zero field), measured at 20 mK, respectively. Shown in the inset of Figure 3**b**, with increasing magnetic field, the system is driven away from the superconducting state, and exhibits a positive linear behavior of the magnetoresistance. This is consistent with the observations shown in Figure 2a, where a positive linear magnetoresistance is also seen in the anomalous metal phase. Importantly, while the longitudinal resistivity becomes non-zero, the Hall resistivity, plotted in Figure 3c and the lower inset, is still vanishing within a much wider range (in the case shown, more than 10 times wider). Consistent with previous observations in $InO_x$, $TaN_x$ and YBCO (ref. 20,24), this vanishing Hall resistivity is an indication of the anomalous metal with particle-hole symmetry. Transverse conductivity $\sigma_{xy}$ as shown in the upper inset is also suppressed approaching the anomalous metal regime. Within the anomalous metal regime showing zero Hall effect, the magnetoresistance develops from linear behavior to exponential behavior (linear in semi-logarithmic conductivity Figure 3**b**, dash line), similar to that observed in MoGe, and $TaN_x$ (ref. 12,24). Interestingly, the onset of non-zero Hall resistance roughly corresponds to the magnetic field where the magnetoresistance/magnetoconductivity deviates from exponential behavior. Higher than this field scale, the system goes into a state with finite Hall resistance (Figure 3**c**), but it is still lower than the Drude value, indicating particle-hole symmetry breaking. We denote this regime as the vortex liquid state[24], indicating delocalized vortices. In the vortex liquid regime, the magnetoresistance follows a linear increasing behavior[12,14,24], dominated by activated flux-flow[27–29]. Further increasing the magnetic field brings the Hall resistivity to a typical Drude linear behavior, reflecting the electron density in the system. Meanwhile, the Cooper pairs are broken and the magnetoresistance deviates from the linear increase and eventually saturates to the Drude value. The remaining reduction of resistivity in this regime comes from the fluctuations of pairing well described by Aslamasov and Larkin[30].

**Phase diagram**

By extracting the transition (superconductor-anomalous metal) and crossover (anomalous metal-vortex liquid and vortex liquid-Drude metal) boundaries from *R-B* curves as a function of *T* (Figure 3**d**), we can map out a phase diagram spanned by *T* and *B*, as shown in the *y-z* plane in Figure 4. Four distinct ground states within our gate and field ranges are identified, in which the boundary between superconductor and anomalous metal represents a phase transition. Carrying out the same analysis to *R-B* curves taken at different $V_G$ and *T* = 20 mK, a field- and gate-tuned phase diagram can also be obtained. Unifying all three phase diagrams, including the one extracted from *R-T* curves (i.e. Figure 1**e**), an experimentally defined global phase diagram is presented in the ($V_G$, *B*, *T*) parameter space. It is worth noting that the transition/crossover points on different planes are obtained from different data sets and determined with different criteria, yet they show high consistency with each other.

**Discussion**

We here discuss the assignments of different regimes in the global phase diagram. The fact that transition and crossover lines coincide on all three axes suggests that the anomalous metal phases on the $V_G$-*B*, *B-T*, and $V_G$-*T* planes have a common physical origin and should be accounted for in one model[1]. This could be seen from the qualitatively identical positive linear magnetoresistance while tuning the system from superconductor to the anomalous metal state with all of the gate, field, and temperature parameters. While thermally activated vortex flow regime on the *B-T* plane is well identified, we see that this regime extrapolates to zero temperature, indicating quantum fluctuations could as well involve vortex motion. This suggests that on the $V_G$-*B* plane, the same vortex liquid state can be identified. On the $V_G$-*T* plane with the absence of magnetic field, topological vortices and anti-vortices play a similar role as external-field-induced vortices, the motion of which could also give rise to a vortex liquid state.

We further discuss the possible role of disorder in this system[5]. Cation interdiffusion[25] at the interface introduces disorder as scattering centers for electrons in the Drude metal state. With the onset of pairing, although the Ginzburg-Landau coherence length, i.e., the size of Cooper pairs (~ 100 nm), is also much larger than disorder length scales (~ 1 nm), emergent inhomogeneity of superconducting puddles occurs as shown by scanning probe measurements[31], consistent with theoretical and numerical predictions indicating that microscopic disorder in 2D naturally gives rise to phase separation[32–34]. Inter-puddle Josephson coupling is weak in the vortex liquid regime such that vortices and anti-vortices are itinerant. In the anomalous metal regime, however, such coupling is enhanced, realizing particle-hole symmetry, yet strong phase fluctuations[35] among different puddles still impose dissipations in transport. Global 2D superconductivity is reached only when the coupling is adequately strong that global phase coherence is achieved.

In conclusion, by unifying the physical picture of the distinct gating, magnetic field, and temperature axes, we show that the anomalous metals are remarkably universal in character: both gate and field tuned samples show quite similar metallic behavior, the hallmark of both being the emergent particle-hole symmetric response, presumably due to pronounced superconducting fluctuations. This should be contrasted with the observation of superconductor-insulator transitions where field tuned and disorder tuned transitions appear to exhibit distinct phenomena[36]. The robustness of the anomalous metallic phases suggests that quantum fluctuations associated with the superconducting order parameter manifest themselves in similar ways regardless of the presence or absence of time-reversal symmetry.

## Methods

**Sample.** Ultraviolet photolithography was used to define an AlO$_x$ hard mask with a Hall-bar pattern (channel width 400 μm) on the SrTiO$_3$ substrate. The 9 unit-cell epitaxial LaAlO$_3$ was grown with pulsed laser deposition at 775 ºC and $1 \times 10^{-5}$ Torr of O$_2$, after pre-annealing at 900 ºC in $5 \times 10^{-6}$ Torr of O$_2$. After growth, the sample was post-annealed at 590 ºC for 3 hours at 1 bar of O$_2$. The gold gate electrode was deposited by electron beam evaporation. The gate pattern is defined by photolithography with wet etching. Electrical contacts to the electrodes were made with ultrasonic Al wire-bonding.

**Magnetotransport measurements.** DC voltages are applied to the gate with respect to the channel. Resistivity measurements were performed using a 100 nA AC current with frequency lower than 200 Hz. Two-wire resistance range from about 0.5 kΩ to 3 kΩ. Strong nonlinear Hall effects are observed at $T$ = 5 K. To correctly estimate the total mobile carriers from the Hall effect, we used the high field limit of the slope (between 13 T and 14 T) of the Hall resistivity. All Hall measurements were obtained from the anti-symmetrized component of the transverse voltage. Low temperature measurements were performed in a dilution refrigerator with a 10-15 mK base temperature measured by a Ruthenium oxide thermometer mounted on the sample space (the temperature quoted in the text and figures). Measurements are performed with low-pass π filters and gold-titanium-plated sapphire plate mounted on the mixing chamber for thermal anchoring of electrons. The cold π filters are individually connected on all wiring leads to the sample. The gold top gate covering the channel and the channel leads on the Hall-bar pattern positioned 9 unit-cell (< 4 nm) away from the conductive interface effectively acts as low-pass RC filters for all (current and voltage) electrical leads and the channel itself with capacitance of ~ 10 nF and cut-off frequencies ~ 100 kHz. To test the filtering effect of the top-gate capacitive configuration, and as a conservative test of the base electron temperature via a high impedance tunnel barrier, we have performed planar tunneling spectroscopy measurements on Ag/LaAlO$_3$/Nb:STO$_3$ samples (without the cold π filters or sapphire-plate anchors)[37,38]. The electronic temperatures obtained by fitting the superconducting gap spectra are shown proportional to thermometer temperature down to below 50 mK.

**Data Availability**

The data that support the findings of this study are available from the corresponding author upon reasonable request.

**Acknowledgements**

We thank S. A. Kivelson and A. Kapitulnik for helpful discussions, and Jiachen Yu for experimental assistance. This work was supported by the Department of Energy, Office of Basic Energy Sciences, Division of Materials Sciences and Engineering, under contract DE-AC02-76SF00515. The dilution refrigerator and associated electronics was acquired by the Gordon and Betty Moore Foundation's EPiQS Initiative through Grant GBMF4415 and GBMF9072.

**Author Contribution**

Z.C. and H.Y.H conceived the experiment. Z.C. synthesized the LaAlO$_3$/SrTiO$_3$ interface, fabricated the devices, performed dilution refrigerator measurements, and analyzed the data.

B.Y.W., A.G.S., H.Y., and Y.H. assisted in experiment implementation. S.R. provided theoretical analysis. Z.C., S.R., and H.Y.H wrote the paper with contributions from all co-authors.

**Competing Interests**

The authors declare that there are no competing interests.

15. Lu, J. M. *et al.* Evidence for two-dimensional Ising superconductivity in gated $MoS_2$. *Science* **350**, 1353–1357 (2015).

16. Li, L. *et al.* Anomalous quantum metal in a 2D crystalline superconductor with electronic phase nonuniformity. *Nano Lett.* **19**, 4126–4133 (2019).

17. Sajadi, E. *et al.* Gate-induced superconductivity in a monolayer topological insulator. *Science* **362**, 922–925 (2018).

18. Fatemi, V. *et al.* Electrically tunable low-density superconductivity in a monolayer topological insulator. *Science* **362**, 926–929 (2018).

19. Garcia-Barriocanal, J. *et al.* Electronically driven superconductor-insulator transition in electrostatically doped $La_2CuO_{4+\delta}$ thin films. *Phys. Rev. B* **87**, 024509 (2013).

20. Yang, C. *et al.* Intermediate bosonic metallic state in the superconductor-insulator transition. *Science* **366**, 1505–1509 (2019).

21. Eley, S., Gopalakrishnan, S., Goldbart, P. M. & Mason, N. Approaching zero-temperature metallic states in mesoscopic superconductor-normal-superconductor arrays. *Nat. Phys.* **8**, 59–62 (2012).

22. Han, Z. *et al.* Collapse of superconductivity in a hybrid tin–graphene Josephson junction array. *Nat. Phys.* **10**, 380–386 (2014).

23. Bøttcher, C. G. L. *et al.* Superconducting, insulating and anomalous metallic regimes in a gated two-dimensional semiconductor–superconductor array. *Nat. Phys.* 14, 1138–1144 (2018).

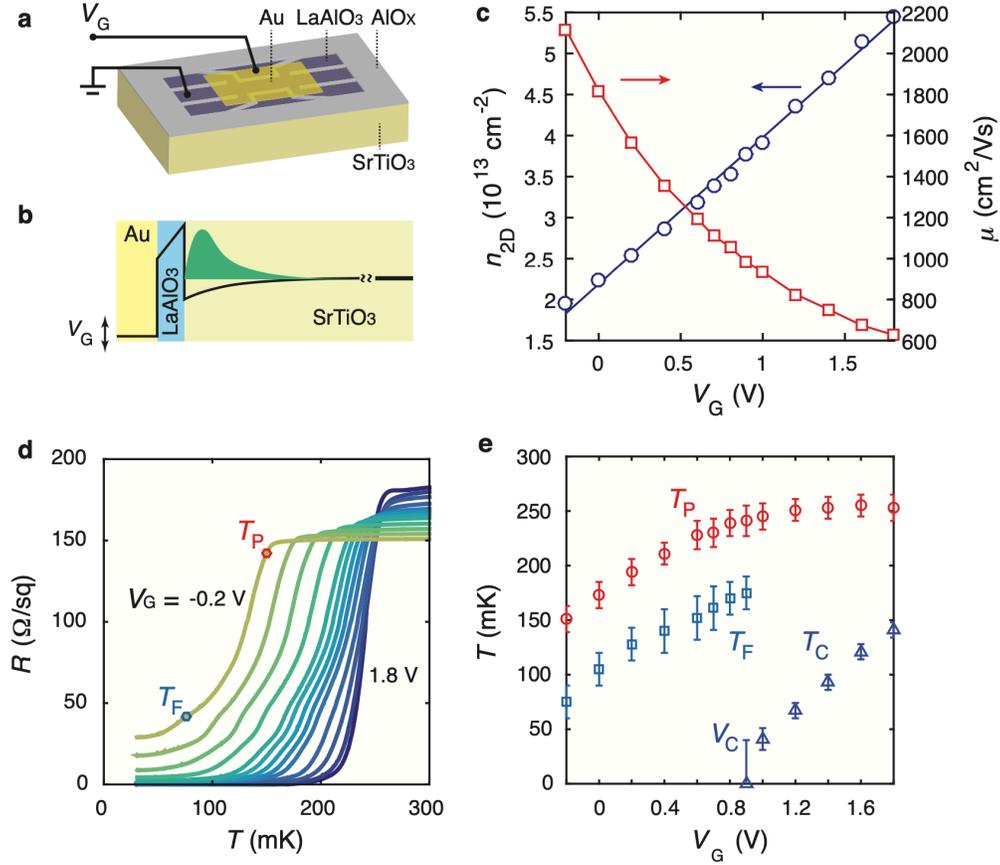

**Figure 1 | Resistivity-versus-temperature (*R-T*) characterization of the gated LaAlO$_3$/SrTiO$_3$ interface. (a)** Schematic of a Au top-gated LaAlO$_3$/SrTiO$_3$ device. 9 unit cells of epitaxial LaAlO$_3$ is grown by pulsed laser deposition on top of SrTiO$_3$ substrate with pre-patterned AlO$x$ hard mask. Channel width is 400 micrometers. **(b)** Schematic diagram for the electrons confined at the interface with density tunable by the top gold gate. The black solid line represents the conduction band bottom. The green shade represents the electron density distribution of the confined electrons. **(c)** Gate modulation of carrier density (circles) and mobility (squares) at *T* = 5 K. The blue solid line is a linear fit to the circles. Red solid curve is a guide to the eyes. **(d)** *R-T* curves as $V_G$ is tuned from – 0.2 V to 1.8 V. Red and blue circles show examples on one curve for the characteristic temperature scales $T_P$ and $T_F$ (extracted from the peaks of the second derivatives -$d^2R/dT^2$), respectively. **(e)** $T_P$, $T_F$, and $T_C$ as functions of $V_G$. Navy triangles, blue squares, and red circles represent $T_C$, $T_F$, and $T_P$, respectively. The critical voltage $V_C$ is determined by the intersect when extrapolating $T_C$ to zero. Errorbars represent confidence intervals. All data in this figure are from Sample A.

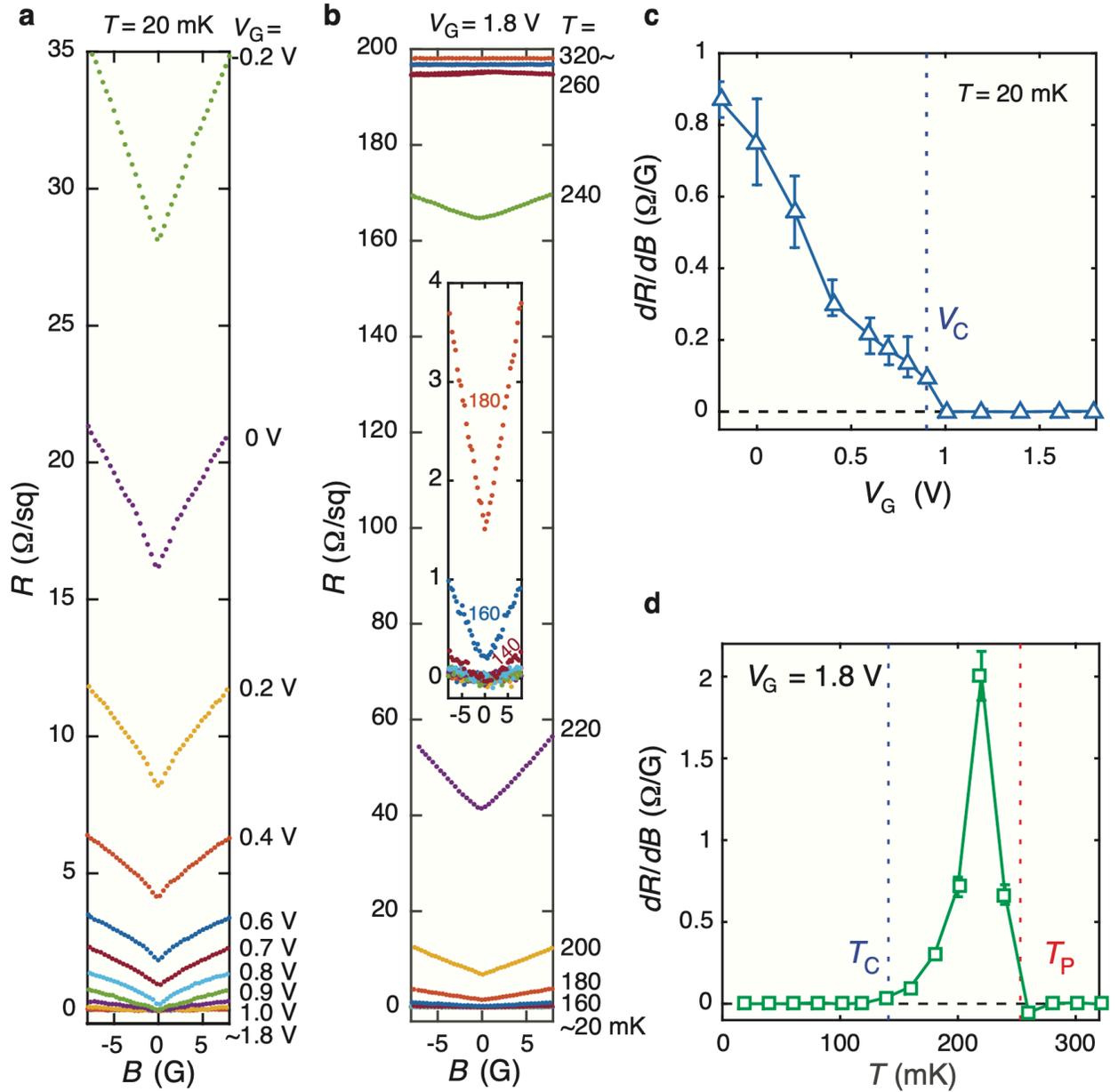

**Figure 2 | Magnetoresistance as a probe of quantum and thermal phase transitions. (a)(b),** Magnetoresistance in the vicinity of zero magnetic field, with varying gate voltage and fixed $T = $ 20 mK and varying temperature and fixed VG = 1.8 V, respectively. Inset of (**b**) is a zoom-in for temperature below 200 mK. (**c**)(**d**), $dR/dB$ near $B = 0$ as a function of $V_G$ and $T$, calculated from **a** and **b**, respectively. Horizontal dashed lines show zero. Vertical dotted lines correspond to $V_C$, $T_C$ and $T_P$, determined from Figure 1. Errorbars represent confidence intervals. All data in this figure are from Sample A.

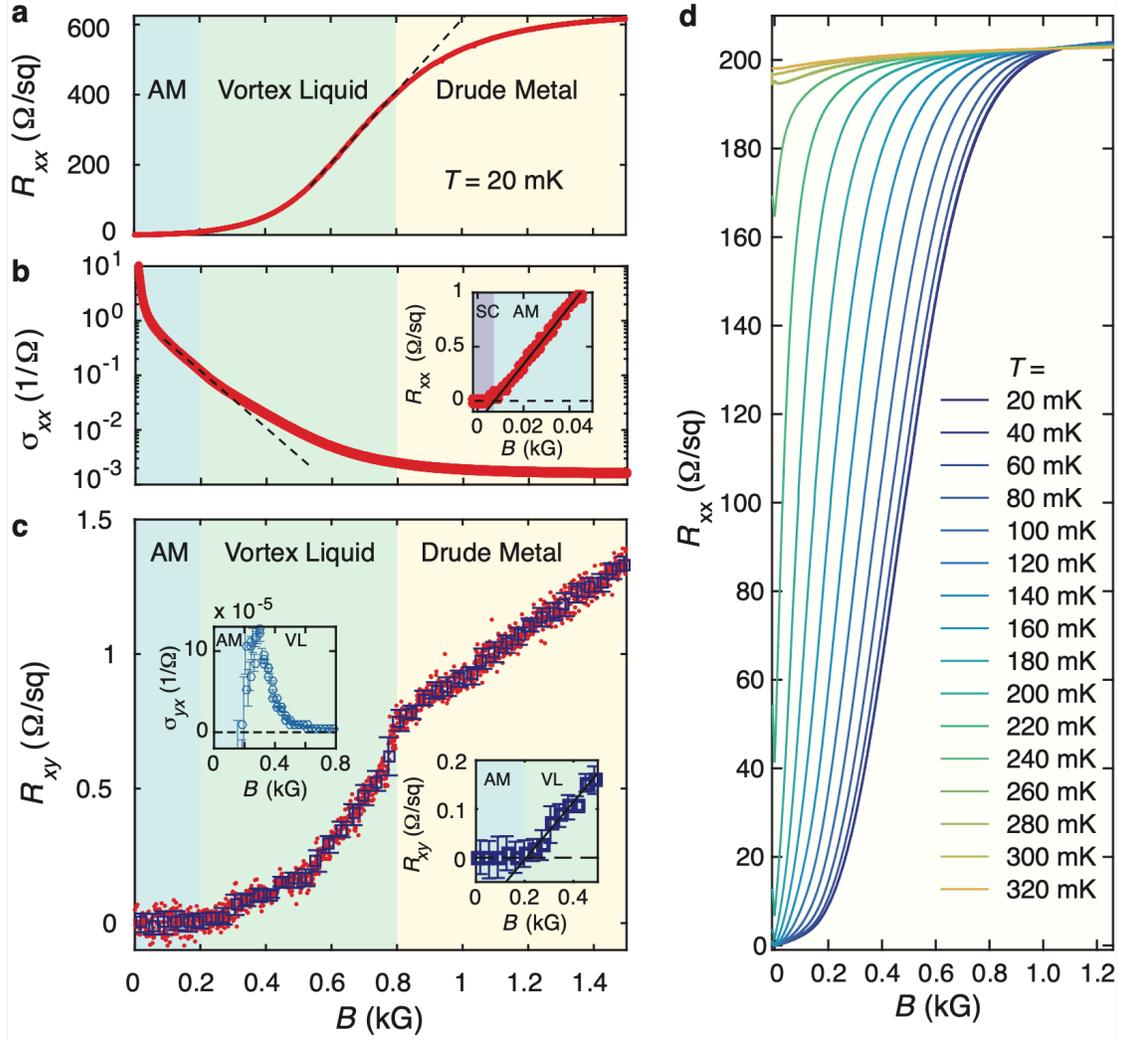

**Figure 3 | Magnetic field tuned superconductor-metal transition. a** and **b,** Resistivity-versus-magnetic field (*R-B*) and conductivity-versus-magnetic field curves measured at *T* = 20 mK and $V_G$ = 1.8 V in sample B, respectively. The dashed lines are an exponential fit for low field data and a linear fit for the intermediate field data in **a** and **b**, respectively. The inset of **b** is a magnification of the low field data in linear scale to show the small superconducting regime. SC: superconductor. AM: anomalous metal. **c,** Hall resistivity as a function of magnetic field at *T* = 20 mK and $V_G$ = 1.8 V in sample B. Red dots are raw data. Blue squares are moving averages of the raw data. The lower inset is a magnification of the low field data to show the onset of non-zero Hall resistivity. The upper inset shows the transverse conductivity $\sigma_{yx}$ as a function of magnetic field. VL: vortex liquid. Errorbars represent standard deviations. **d,** *R-B* curves at different temperatures of sample A.

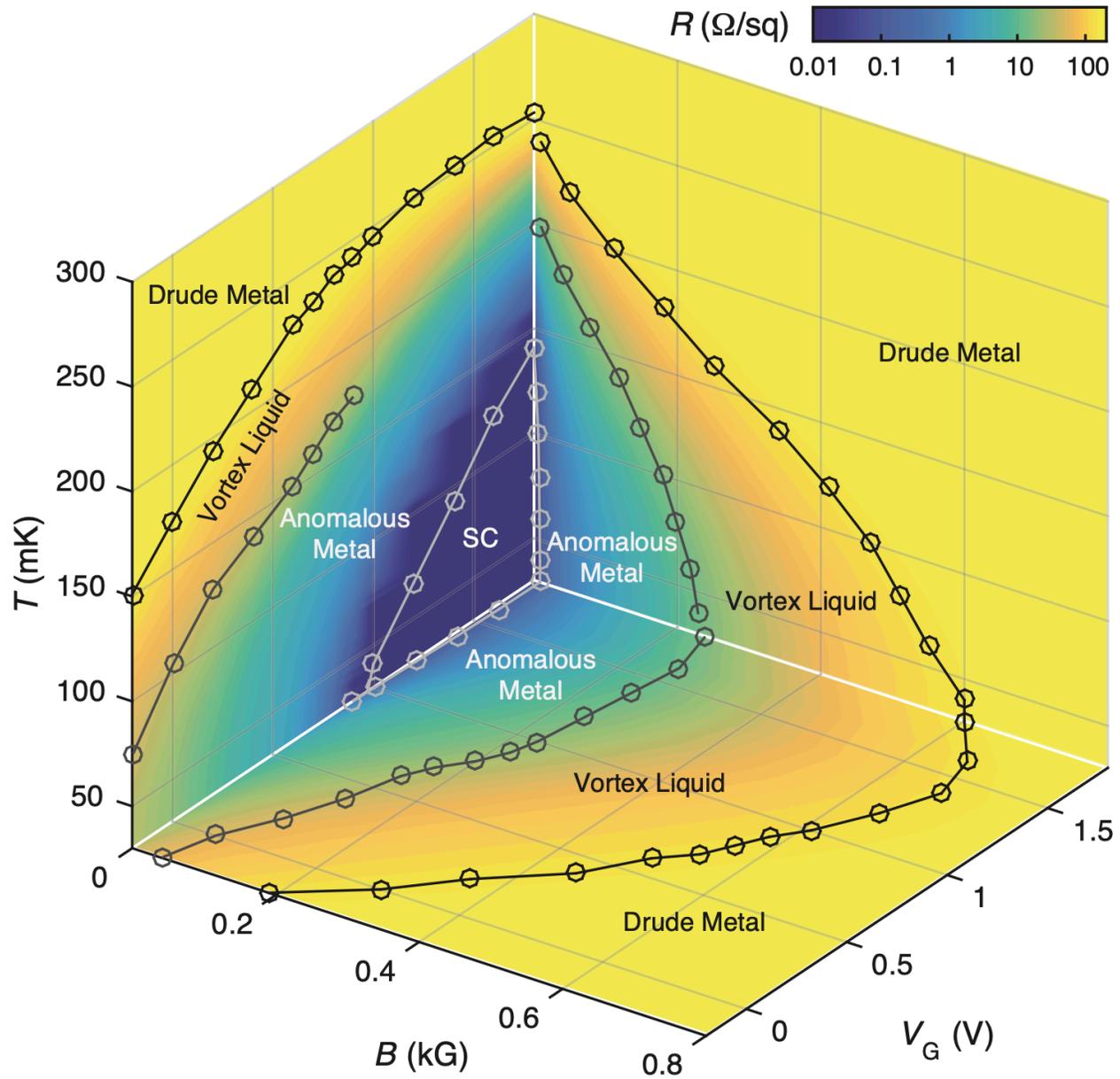

**Figure 4 | Global phase diagram in the ($V_G$, $B$, $T$) parameter space.** Transitions/crossovers points are defined by methods discussed in the main text. Note that on the ($V_G$, $B$) and ($B$, $T$) planes, the superconducting (SC) regime has a very small experimental width along the $B$ axis. Color scale in the background shows the interpolated resistivity. All data in this figure are from Sample A.

20